\documentstyle[11pt,epsf]{article}
\newtheorem{theorem}{Theorem}

\newcommand {\dfn} {\stackrel{\Delta} {=}}
\newcommand {\exe} {\stackrel{\cdot} {=}}
\newcommand {\gexe} {\stackrel{\cdot} {\ge}}
\newcommand {\lexe} {\stackrel{\cdot} {\le}}

\newcommand{\gea}{\stackrel{\mbox{\tiny (a)}}{\ge}}
\newcommand{\geb}{\stackrel{\mbox{\tiny (b)}}{\ge}}
\newcommand{\gec}{\stackrel{\mbox{\tiny (c)}}{\ge}}
\newcommand{\ged}{\stackrel{\mbox{\tiny (d)}}{\ge}}
\newcommand{\gee}{\stackrel{\mbox{\tiny (e)}}{\ge}}
\newcommand{\gef}{\stackrel{\mbox{\tiny (f)}}{\ge}}
\newcommand {\reals} {{\rm I\!R}}

\newcommand {\bx} {\mbox{\boldmath $x$}}
\newcommand {\by} {\mbox{\boldmath $y$}}
\newcommand {\bz} {\mbox{\boldmath $z$}}

\newcommand {\bE} {\mbox{\boldmath $E$}}

\newcommand {\bY} {\mbox{\boldmath $Y$}}
\newcommand {\bZ} {\mbox{\boldmath $Z$}}

\newcommand{\calC}{{\cal C}}

\newcommand{\calE}{{\cal E}}
\newcommand{\calF}{{\cal F}}

\newcommand{\calH}{{\cal H}}
\newcommand{\calI}{{\cal I}}

\newcommand{\calO}{{\cal O}}

\newcommand{\calS}{{\cal S}}
\newcommand{\calT}{{\cal T}}

\begin{document}
\thispagestyle{empty}
\title{Trade-offs Between Weak--Noise Estimation Performance and Outage Exponents in
Nonlinear Modulation}
\author{Neri Merhav}
\date{}
\maketitle

\begin{center}
The Andrew \& Erna Viterbi Faculty of Electrical Engineering\\
Technion - Israel Institute of Technology \\
Technion City, Haifa 32000, ISRAEL \\
E--mail: {\tt merhav@ee.technion.ac.il}\\
\end{center}
\vspace{1.5\baselineskip}
\setlength{\baselineskip}{1.5\baselineskip}

\begin{abstract}
We focus on the problem of modulating a parameter onto a power--limited signal
transmitted over a discrete--time Gaussian channel
and estimating this parameter at the receiver. Considering the well--known
threshold effect in non--linear modulation systems, 
our approach is the following: instead
of deriving upper and lower bounds
on the total estimation error, which weigh both weak--noise errors and
anomalous errors beyond the threshold, we separate the two kinds of errors. In
particular, we derive upper and lower bounds on the best achievable trade-off between the
exponential decay rate of the weak--noise expected error cost 
and the exponential decay rate of the 
probability of the anomalous error event, also referred
to as the {\it outage event}. This outage event is left to be defined as part of
the communication system design problem. Our achievability scheme, which is
based on lattice codes, meets the lower bound at the high signal--to--noise
(SNR) limit and for
a certain range of trade--offs between the weak--noise error cost and the
outage exponent.\\

\end{abstract}

\section*{I. Introduction}

Consider the problem of conveying a real--valued parameter $u$ by
means of
$n$ uses of the additive white Gaussian noise (AWGN) channel,
\begin{equation}
y_i=x_i+z_i,~~~~~~~i=1,2,\ldots,n,
\end{equation}
where $x_i$ is the $i$--th component of a channel input vector,
$\bx=(x_1,x_2,\ldots,x_n)=f_n(u)$, that depends on
the parameter $u$, and that is subjected to a power constraint,
$\|\bx\|^2\le nP$, $\{z_i\}$ are independent, zero--mean, Gaussian
random variables with variance $\sigma^2$, and $y_i$ is the $i$--th component
of the channel output vector, $\by=(y_1,y_2,\ldots,y_n)$.
In a nutshell, our central interest, 
in this work, is to address the following question: how well can
one estimate $u$
based on $\by$ when one is allowed to optimize, not only the estimator, but
also the modulator, that is, the function $f_n(\cdot)$ that
maps $u$ into a channel
input vector? How fast does the estimation error decay as a function of $n$
when the best modulator and estimator are used?

In principle, this problem, which is the discrete--time analogue of the
classical problem of ``waveform communication'' (in the terminology of
\cite[Chap.\ 8]{WJ65}),
can be viewed both from the information--theoretic and the
estimation--theoretic perspectives. In the former, this is an instance of the
joint source--channel coding problem (see, e.g., \cite{KGT17} and references
therein), where a single source symbol $u$ is transmitted by using the channel
$n$ times, and it also falls within the framework of Shannon--Kotel'nikov
mappings (see, e.g., \cite{Floor08}, \cite{Hekland07}, \cite{HFR09},
\cite{KR07} and
references therein). From the estimation--theoretic point of view, every given
modulator induces a certain parametric family of conditional densities of
$\by$ given $u$, and estimation theory provides a plethora of both Bayesian
and non--Bayesian lower
bounds (depending on whether $u$ is a deterministic parameter or a random
variable) on the estimation performance, as well as useful estimators that
perform well, e.g., the maximum likelihood (ML) estimator in the non--Bayesian
case, the maximum
a--posteriori (MAP) estimator in the Bayesian case, and others.

A well known inherent 
problem in the design of non--linear modulators is the {\it threshold effect}
(see, e.g., \cite[Chap.\ 8]{WJ65}). The threshold effect means a sharp
transition between two modes of behavior when the signal--to--noise ratio (SNR)
crosses a certain critical value. For high SNR, a.k.a.\ the {\it weak--noise}
regime, the estimation error behaves 
similarly as the one attained for a linear modulator, where it roughly achieves
the Cram\'er--Rao lower bound (CRLB). But beyond a certain noise level, the
estimation performance completely breaks down rather abruptly. As explained in
\cite[Chap.\ 8]{WJ65}, for a given non--linear modulator, one can identify a
certain anomaly event, or {\it outage event}, whose probability becomes (quite
abruptly) appreciably large as the threshold noise level is crossed. 

The literature on the problem of non--linear modulation and estimation
contains plenty of results in the form of performance bounds (see, e.g.,
\cite{Burnashev84}, \cite{Burnashev85}, \cite{Cohn70} and many references
therein), in which there is no distinction between weak--noise errors and
anomalous errors, i.e., both types of errors weigh in the
evaluation of the total mean square error (MSE). A little thought,
however, suggests that it makes a lot of sense to separate the two kinds of
errors because estimation in the presence of an outage event is rather
meaningless. Indeed, separate treatments of the two kinds of errors appear
already in \cite[pp.\ 661--674]{WJ65}, but not quite in a formal manner.
A more systematic approach in this direction, of separating weak--noise errors
from anomalous errors, appears in \cite[Section IV.A]{KGT17}, where the problem
was posed in terms of designing a communication system, along with a
definition of an outage event (albeit with a different motivation in mind), 
with the target of minimizing the MSE given
that the outage event has not occurred, subject to the constraint
that the outage probability would not exceed a given (small) constant.
It was shown in \cite{KGT17}, that the data processing lower bound is asymptotically
achieved, in this setup, by a conceptually simple scheme, that first quantizes the
parameter and then maps the set of quantized parameter values into a good
digital channel code for the Gaussian channel. The receiver first decodes the
digital message and then maps it back to the corresponding quantized parameter
value. The outage event is then the error event in the digital part and then
the weak--noise MSE is simply the quantization error. 
The missing link in this result,
however, is that the data processing lower bound is not quite compatible to
this setting, as it corresponds to a situation where there is no freedom to
allow an outage event.

In this paper, we sharpen the approach of \cite{KGT17} in two ways: firstly, both
in the lower bound and in the upper bound, we allow an outage
event, whose definition is subject to optimization, depending on the
communication system itself. Therefore, the 
lower bound and the upper bound refer to the same setting.
Secondly, we somewhat refine the ``resolution'' in the quantitative aspect of
the problem being addressed:
instead of constraining the outage probability to be upper bounded by a small
constant, we impose the constraint that the outage probability would not
exceed a given exponential function of $n$, that is, $e^{-\lambda n}$ for some
prescribed positive constant $\lambda$. Under such a constraint, we seek the
fastest possible decay rate of the MSE (as a function of $\lambda$ and the
SNR, $P/\sigma^2$), 
or more generally, the expectation of
an arbitrary symmetric, convex function of the estimation error.
More precisely, we derive an upper bound and a lower bound, which coincide in
the limit of high SNR for a certain range of values of $\lambda$. 
Our proposed achievability scheme is based on quantization and modulation, as described
before, but since the optimal outage event turns out to be the complement of 
a sphere in the
space of noise vectors (due to the Gaussianity of the noise), this suggests
the use of good {\it lattice codes}, whose Voronoi cells can indeed be shaped arbitrarily
closely to $n$--dimensional spheres \cite[Chap.\ 7]{Zamir14}, for large $n$. 

The paper is organized as follows. In Section II, we define notation
conventions and formalize the problem being addressed along with the main assumptions.
In Section III, we provide the converse bound. In Section IV, we
describe the achievability scheme, and finally, in Section V, we summarize and
conclude.

\section*{II. Notation, Problem Formulation and Assumptions}

Consider the following communication system. The transmitter wishes to convey
to the receiver, a
real--valued parameter $u$, taking on values in a finite
interval, say, without essential loss of generality, the interval $[0,1]$. To
this end, the transmitter uses the channel $n$ times, subject to a given power
constraint. The receiver has to estimate $u$ from the $n$ noisy channel outputs.
More precisely, given $u\in [0,1]$, the transmitter sends 
a vector $\bx=f_n(u)\in\reals^n$, whose power is limited by $\|\bx\|^2\le nP$,
where $P$ is the maximum allowed average power per channel use. 
The received vector is
\begin{equation}
\bY=f_n(u)+\bZ,
\end{equation}
where $\bZ\in\reals^n$ is a zero--mean Gaussian noise vector with covariance
matrix $\sigma^2\cdot I$, $I$ being the $n\times n$ identity matrix.
The receiver implements an estimator $\hat{u}=g_n[\by]$ (with $\by$ designating a
realization of the random vector $\bY$) 
of the parameter $u$, where $g_n:\reals^n\to[0,1]$.
For every $u\in[0,1]$, let $\calO_n(u)\subset\reals^n$ designate an event defined in the
space of noise vectors, $\{\bz\}$, which is henceforth referred to as the {\it outage event} (or
the {\it anomalous error event}) given $u$. 

Our general objective is to design a communication
system, defined by a transmitter (which is a modulator), 
$f_n(\cdot)$ that satisfies the aforementioned power constraint, and a receiver (which is
an estimator), $g_n[\cdot]$, along with a family of outage events,
$\calO_n=\{\calO_n(u),~0\le u\le 1\}$, so as to minimize 
\begin{equation}
\label{obj}
\sup_{u\in[0,1]}\bE\left\{\rho(g_n[f_n(u)+\bZ]-u)\bigg|\calO_n^{\mbox{\tiny
c}}(u)\right\} 
\end{equation}
subject to the constraint that
\begin{equation}
\label{constraint}
\sup_{u\in[0,1]}\mbox{Pr}\{\calO_n(u)\}\le e^{-\lambda n},
\end{equation}
where the expectation $\bE\{\cdot\}$ in (\ref{obj}) and the probability
$\mbox{Pr}\{\cdot\}$ in
(\ref{constraint}) are with respect
to (w.r.t.) the randomness of the noise vector $\bZ$. Here,
$\lambda > 0$ is a prescribed constant (independent of $n$), henceforth
referred to as the {\it outage exponent},
$\calO_n^{\mbox{\tiny c}}(u)$ is the complement of $\calO_n(u)$,
and $\rho:\reals\to\reals^+$ is referred to as
the {\it error cost function} (ECF), which is assumed to have the following properties:
(i) symmetry: $\rho(t)=\rho(-t)$, (ii) convexity, (iii) increasing 
monotonicity for $t\ge 0$ (and hence decreasing monotonicity for $t\le 0$), and
(iv) $\rho(0)=0$. 
Let $\calC_n(\lambda)$ denote the class of all families of
outage events, $\{\calO_n\}$, that satisfy (\ref{constraint}).

For certain ECFs, like the squared error, $\rho(t)=t^2$, and more generally,
for $\rho(t)=|t|^\alpha$, $\alpha\ge 1$, it is well known, from classical
communication theory (see, e.g., \cite[Chap.\ 8]{WJ65}), that there exist
modulators, receivers and families of outage events for which  
(\ref{obj}) decays exponentially with $n$. Since the above defined constrained
optimization problem is difficult to solve in a precise closed form, we will adopt the
customary information--theoretic approach of
characterizing the fastest possible exponential decay rate of (\ref{obj})
subject to (\ref{constraint}). More formally, for given sequences of
modulators $F=\{f_n(\cdot)\}_{n\ge 1}$ (all satisfying the power constraint), 
estimators $G=\{g_n[\cdot]\}_{n\ge 1}$, and families
of outage events, $\calO=\{\calO_n\}_{n\ge 1}$ (with
$\calO_n\in\calC_n(\lambda)$), let
\begin{equation}
\label{ecfexponent}
\calE(F,G,\calO)=\liminf_{n\to\infty}\left[-\frac{1}{n}\ln\left(
\sup_{u\in[0,1]}\bE\left\{\rho(g_n[f_n(u)+\bZ]-u)\bigg|\calO_n^{\mbox{\tiny
c}}(u)\right\}\right)\right].
\end{equation}
Our purpose is to derive upper and lower bounds to the best achievable
value of $\calE(F,G,\calO)$ as functions of the outage exponent, $\lambda$,
and the SNR, $\gamma=P/\sigma^2$.
In particular, we derive simple formulas for upper and lower bounds to
the best achievable high--SNR error--cost exponent
$\calE(F,G,\calO)$, denoted $E_{\mbox{\tiny U}}(\lambda,\gamma)$ and 
$E_{\mbox{\tiny L}}(\lambda,\gamma)$, respectively, 
which are asymptotically compatible in the sense that
$\lim_{\gamma\to\infty}[E_{\mbox{\tiny U}}(\lambda,\gamma)-E_{\mbox{\tiny
L}}(\lambda,\gamma)]=0$, for a
certain range of values of the outage exponent, $\lambda$.

To this end, we will make three additional
assumptions, the first one concerns the ECF, the second is about the
modulator, and third one is regarding the outage events.

\begin{itemize}

\item[A.1] For any given constant $c >
0$, we assume that $\rho(e^{-nc})\exe e^{-n\zeta(c)}$, where $\zeta(\cdot)$ 
is some continuous function 
with the property that $c> 0$ implies $\zeta(c) > 0$, and where $\exe$ denotes equivalence in the
exponential scale: the notation $a_n\exe b_n$, for two positive sequences,
$\{a_n\}$ and $\{b_n\}$, means that $\frac{1}{n}\log\frac{a_n}{b_n}\to 0$, as
$n\to\infty$. Moreover, we will assume that $\alpha_n\exe e^{-nc}$ implies
$\rho(\alpha_n)\exe e^{-n\zeta(c)}$.\footnote{The function $\zeta(\cdot)$ is, of course, induced by
the function $\rho$. For example, if $\rho(t)=|t|^\alpha$, then obviously,
$\zeta(c)=\alpha c$.
As a side remark, note that the function $\rho(\cdot)$ is allowed to depend on
$n$.}

\item[A.2]
Denoting $M_n=e^{nR
(\lambda,\gamma)}$ (with $R(\lambda,\gamma)$ to be defined
in Section III), consider the partition of the unit interval into $M_n$
non--overlapping sub--intervals of length $1/M_n$. Then, it will be assumed
that for all sufficiently large $n$, the number, $M_n^{\mbox{\tiny c}}$,
of sub--intervals in which $f_n(\cdot)$ is continuous, is of the exponential
order of $M_n$, i.e., $M_n^{\mbox{\tiny c}}\exe e^{nR
(\lambda,\gamma)}$. Also, in each sub--interval $[a,b]$ of
continuity, the integral $\int_{f_n(a)}^{f_n(b)}\|\mbox{d}f_n(u)\|$ exists and
is finite.\footnote{Note that continuity alone does not guarantee that
this first order variation integral is finite. A simple counter--example is
Brownian motion.}
Henceforth, we define $\calF_n(P)$ as the class of
modulators that satisfy these conditions, in addition to the power constraint,
$\|f_n(u)\|^2\le nP$.

\item[A.3]
If, for a certain vector $\bz\in \calO_n(u)$, one of the
components vanishes, then upon replacing this component by any non--zero number,
the resulting vector remains in $\calO_n(u)$.
In addition, it is assumed that as noise variance tends to zero, the covering
radius of $\calO_n^{\mbox{\tiny c}}(u)$ tends to zero as well.
\footnote{The first point is reasonable 
since the new vector designates ``stronger noise'' than the original one.
The second requirement also makes sense, because as the noise becomes weaker,
the non--outage events may shrink too (and in all directions on $\reals^n$), 
in order to improve the weak--noise
estimation performance without violating the outage constraint.}

\end{itemize}

The lower bound,
$E_{\mbox{\tiny L}}(\lambda,\gamma)$, will be achieved by a conceptually simple
achievability scheme, described as follows: we first uniformly quantize the
parameter into a grid of $M=e^{nR}$ ($R > 0$) points, and then map this
set of points into a good rate--$R$ lattice code of dimension $n$. The coding
rate $R$ will be used to control the trade--off between the outage exponent,
$\lambda$, and the error--cost exponent,
$E_{\mbox{\tiny L}}(\lambda,\gamma)$. The reason for the choice of lattice
codes will become apparent from the derivation of the upper bound,
$E_{\mbox{\tiny U}}(\lambda,\gamma)$. 

A few comments are in order concerning the above problem formulation.
\begin{enumerate}

\item The assumption that $u$ takes on values in the unit interval is made
largely for the sake of convenience. The extension to any other finite
interval, $[a,b]$, will be straightforward. Concerning the extension to the
entire real line, the converse part may remain intact, but for the
achievability part, the supremum over $u$ in (\ref{ecfexponent}) would have to
be confined to an interval that grows with $n$ slowly enough, i.e., the interval
$[-A_n,+A_n]$ where $A_n$ grows at a sub--exponential rate.
Anyway, with this formulation, the entire real line is eventually covered in the limit
$n\to\infty$.

\item Here, we adopt the minimax approach, a.k.a.,
the worst--case approach, of minimizing the estimation performance for the
worst--case value of $u$ in $[0,1]$. Owing to the fact that we focus on
exponential error bounds, this is asymptotically equivalent to
the Bayesian setting as long as the prior of $u$ is bounded away from zero and
infinity. This is because given a function $s(u)$ (that does not depend on
$n$), $\int_0^1 p(u)e^{-ns(u)}\mbox{d}u$ is dominated by
$\sup_{u\in[0,1]}e^{-ns(u)}=\exp\{-n\inf_{u\in[0,1]}s(u)\}$. 

\item Since the estimation errors considered are in an exponentially small
scale, it is actually only important how the function $\rho(\cdot)$ behaves in the
vicinity of the origin. For most of the conceivable convex symmetric cost
functions, near the origin, $\rho(t)$ is proportional to $|t|^\alpha$ for some
$\alpha\ge 1$.
Thus, power cost functions are essentially as general as any convex cost
function for our purposes.

\item While this point is well known, we nonetheless feel compelled to emphasize that
the index $i$ of each component $x_i$ of the vector $\bx=f_n(u)$ 
(as well as the those of $\by$ and
$\bz$) need not necessarily designate discrete time. More generally, $\bx$
should be thought of as a vector of coefficients that represent the
transmitted signal as a linear combination of $n$ arbitrary orthonormal basis
functions (in either discrete time or continuous time).
In particular, if these basis functions are taken to be sinusoids (or complex exponentials),
then the index $i$ of each component $x_i$ designates frequency. For
example, the space of time--limited and (approximately)
band--limited signals, of duration $T$ and bandwidth $W$, are well known to be
spanned by $n\approx 2WT$ basis functions for $WT \gg 1$ (see, e.g.,
\cite{Wyner73}). Thus, our
setup can easily be translated into the framework of the continuous--time, band--limited
AWGN channel, with $\sigma^2$ being replaced by the noise spectral density
$N_0/2$, the discrete block--length $n$ will be replaced by the time duration
$T$, and all the exponents should be multiplied by the factor $2W$, owing to
the substitution of $n$ by $2WT$ in all places (see also \cite{WM17}).

\item The achievability part in \cite[Section IV.A]{KGT17},
where the outage probability was kept below a small constant
$\delta > 0$, is clearly
a special case of our problem formulation with $\lambda=\frac{1}{n}\ln\frac{1}{\delta}
\to 0$. The performance of the achievability scheme in \cite{KGT17},
however, was contrasted with the traditional data processing converse bound,
that allows no outage at all, namely, $\calO(u)$ must be an empty set for all
$u$. But this is equivalent to the special case of our framework with $\lambda=\infty$.
This observation displays a considerable 
mismatch between the settings of the achievability and the converse
parts in \cite{KGT17}, when formalized in our framework. As a side remark, we
should point out that in \cite{KGT17}, the Bayesian approach was adopted,
where the parameter $u$ was assumed to be a Gaussian random variable.
\end{enumerate}

\section*{III. Converse Bound}

Given a positive real $\lambda$, let $w(\lambda)$ be the unique solution
$\theta$ to
the equation
\begin{equation}
\label{wlambda}
\theta-\ln(1+\theta)=2\lambda.
\end{equation}
Let us also define
\begin{equation}
R(\lambda,\gamma)=\frac{1}{2}\ln\frac{\gamma}{1+w(\lambda)}
\end{equation}
and
\begin{equation}
\label{eu}
E_{\mbox{\tiny U}}(\lambda,\gamma)=
\zeta[R(\lambda,\gamma)],
\end{equation}
where $\zeta(\cdot)$ is as defined in Assumption A.1, in Section II.
Our first main result is the following.

\begin{theorem}
Under the assumptions of Section II,
for every sequence $F$ of modulators ($f_n\in\calF_n(P)$),
every sequence $G$ of estimators, and every sequence $\calO$ of 
families of outage events, 
\begin{equation}
\limsup_{\gamma\to\infty}[\calE(F,G,\calO)-
E_{\mbox{\tiny U}}(\lambda,\gamma)]\le 0.
\end{equation}
\end{theorem}

Before we prove Theorem 1, a brief informal discussion about 
the plan of the proof is in order.
Consider first the relatively simple case where the modulator, $f_n(\cdot)$, is continuous
across the entire unit interval, $[0,1]$ (the idea can then be extended also to the
case where there are discontinuities). As $u$ exhausts the unit interval,
$\bx=f_n(u)$ draws a curve inside the sphere, $\calS_n(P)\dfn\{\bx:~\|\bx\|^2\le nP\}$.
We refer to this curve as the {\it signal locus} associated with the modulator
$f_n$. The length of the signal locus is given by
\begin{equation}
L(f_n)=\int_{f_n(0)}^{f_n(1)}\|\mbox{d}f_n(u)\|,
\end{equation}
which is an integral that exists according to Assumption A.2.
If, in addition, $f_n(\cdot)$ is differentiable for all $u\in(0,1)$, we may 
write
\begin{equation}
L(f_n)=\int_0^1\|\dot{f}_n(u)\|\mbox{d}u,
\end{equation}
where $\dot{f}_n(u)$ is the vector of derivatives of the components of
$f_n(u)$ w.r.t.\ $u$. Our proof plan is as follows: we first derive 
a lower bound on the estimation performance,
$\bE\{\rho(g_n[f_n(u)+\bZ]-u)|\calO_n^{\mbox{\tiny
c}}(u)\}$, which is a monotonically non--increasing function of $L(f_n)$.
Then, we derive an upper
bound, $L_n^*$, on $L(f_n)$, that 
must apply (at least in the exponential scale) 
to every $f_n\in\calF_n(P)$ in the high SNR
regime. This is carried out using certain volume considerations, a.k.a.\ 
``tube packing'' (in analogy to sphere packing in channel coding),
which will also imply that for all $u$, the best choice of the outage
event, $\calO_n(u)$, would be the complement of the sphere, centered at the
origin, whose radius is exactly large
enough to satisfy the outage constraint.
Finally, upon substituting $L(f_n)$ by $L_n^*$ in the 
lower bound to the estimation performance, we
arrive at our ultimate lower bound that no longer depends on
the specific modulator, $f_n$. The proof of Theorem 1 is then completed by
assessing the exponential rate of the lower bound in the high SNR regime.

We should point that generally speaking, the derivations in \cite[Chap.\
8]{WJ65} are guided by essentially the same considerations, but there are some important
differences. The main difference is that in \cite{WJ65}, there is no really 
a derivation of a lower bound on the estimation performance, but a rather
informal argument that concerns the weak--noise MSE of the ML estimator. Now, 
the weak--noise MSE of the ML estimator
is inversely proportional to the energy of the
signal derivative, $\|\dot{f}_n(u)\|^2$.
Since this $L_2$ norm is not related to the length, $L(f_n)$ in a unique
manner, a minimax consideration is invoked in \cite[p.\ 620]{WJ65} in order to
convince the reader that it is best to confine attention to modulators for
which $\|\dot{f}_n(u)\|$ is identical to a constant $S$ (independent of $u$)
and then $L(f_n)=S$ and the energy of the derivative is simply $S^2$, and so, they now become
related in simple manner. There are at least three weaknesses in this kind of argument: (i) it
assumes that $f_n$ is differentiable, (ii) it applies to the ML estimator, but 
it is not a lower bound for an arbitrary estimator, and (iii) it is not
quite clear that the confinement to modulators with the property 
$\|\dot{f}_n(u)\|\equiv S$,
does not harm the trade--off with the outage performance. Our derivation, on the other
hand, does not suffer from these weaknesses, because, as said, the estimation performance is
bounded directly in terms of the signal locus length, not in terms of the energy of the
derivative, which may not even exist for a general modulator.

\noindent
{\bf Proof of Theorem 1.}
Let $f_n$, $g_n$ and $\calO_n$ be given and let Assumptions A.1--A.3 be
satisfied. Let $M$ be
a positive integer, which will be specified later. For simplicity, we assume
first that $f_n(\cdot)$ is continuous along the entire interval (and describe the
modifications needed when this is not the case, in short discussion in the
sequel). Then, denoting the density of noise vector, $\bZ$, by $p(\bz)$, and the
indicator function of the event $\calO_n^{\mbox{\tiny c}}(u)$, by
$\calI\{\calO_n^{\mbox{\tiny c}}(u)\}$, we have:
\begin{eqnarray}
\label{lowerbound}
& &\sup_{u\in[0,1]}\bE\left\{\rho(g_n[\bY]-u)\bigg|\calO_n^{\mbox{\tiny
c}}(u)\right\}\nonumber\\
&\ge&\frac{1}{2M}\sum_{i=0}^{M-1}
\left[\bE\left(\rho(g_n[\bY]-u_i)\cdot\calI\{\calO_n^{\mbox{\tiny
c}}(u_i)\}\right)+
\bE\left(\rho(g_n[\bY]-u_{i+1})\cdot\calI\{\calO_n^{\mbox{\tiny
c}}(u_{i+1})\}\right)\right]
\nonumber\\
&\gea&\frac{1}{2M}\sum_{i=0}^{M-1}
\int_{\calO_n^{\mbox{\tiny c}}(u_i)\cap\calO_n^{\mbox{\tiny c}}(u_{i+1}}
\left[p(\by-f_n(u_i))\rho(g_n[\by]-u_i)+\right.\nonumber\\
& &\left. p(\by-f_n(u_{i+1}))\rho(u-g_n[\by])\right]
\mbox{d}\by\nonumber\\
&\ge&\frac{1}{M}\sum_{i=0}^{M-1}
\int_{\calO_n^{\mbox{\tiny c}}(u_i)\cap\calO_n^{\mbox{\tiny c}}(u_{i+1}}
\min\{p(\by-f_n(u_i)),p(\by-f_n(u_{i+1}))\}\times\nonumber\\
& &\left[\frac{1}{2}\rho(g_n[\by]-u_i)+
\frac{1}{2}\rho(u_{i+1}-g_n[\by])\right]
\mbox{d}\by\nonumber\\
&\geb&\frac{1}{M}\sum_{i=0}^{M-1}
\int_{\calO_n^{\mbox{\tiny c}}(u_i)\cap\calO_n^{\mbox{\tiny c}}(u_{i+1})}
\min\{p(\by-f_n(u_i)),p(\by-f_n(u_{i+1}))\}\times\nonumber\\
& &\rho\left(\frac{g_n[\by]-u_i}{2}+
\frac{u_{i+1}-g_n[\by]}{2}\right)
\mbox{d}\by\nonumber\\
&=&\frac{1}{M}\sum_{i=0}^{M-1}\rho\left(\frac{u_{i+1}-u_i}{2}\right)
\int_{\calO_n^{\mbox{\tiny c}}(u_i)\cap\calO_n^{\mbox{\tiny c}}(u_{i+1})}
\min\{p(\by-f_n(u_i)),p(\by-f_n(u_{i+1}))\}
\mbox{d}\by\nonumber\\
&=&\rho\left(\frac{1}{2M}\right)\cdot\frac{1}{M}\sum_{i=0}^{M-1}
\int_{\calO_n^{\mbox{\tiny c}}(u_i)\cap\calO_n^{\mbox{\tiny c}}(u_{i+1})}
\min\{p(\by-f_n(u_i)),p(\by-f_n(u_{i+1}))\}\mbox{d}\by\nonumber\\
&=&2\rho\left(\frac{1}{2M}\right)\cdot\frac{1}{M}\sum_{i=0}^{M-1}
\left[\frac{1}{2}\int_{\reals^n}\min\{p(\by-f_n(u_i)),p(\by-f_n(u_{i+1}))\}\mbox{d}\by
-\right.\nonumber\\
& &\left.-\frac{1}{2}\int_{\calO_n(u_i)\cup\calO_n(u_{i+1})}
\min\{p(\by-f_n(u_i)),p(\by-f_n(u_{i+1}))\}\mbox{d}\by\right]\nonumber\\
&\gec&2\rho\left(\frac{1}{2M}\right)\cdot\frac{1}{M}\sum_{i=0}^{M-1}
\left[Q\left(\frac{\|f_n(u_{i+1})-f_n(u_i)\|}{2\sigma}\right)-\right.\nonumber\\
& &\left.\frac{1}{2}\int_{\calO_n(u_i)}p(\by-f_n(u_i))\mbox{d}\by-
\frac{1}{2}\int_{\calO_n(u_{i+1})}p(\by-f_n(u_{i+1}))\mbox{d}\by\right]\nonumber\\
&\ged&2\rho\left(\frac{1}{2M}\right)\cdot\frac{1}{M}\sum_{i=0}^{M-1}
\left[Q\left(\frac{\|f_n(u_{i+1})-f_n(u_i)\|}{2\sigma}\right)-e^{-\lambda
n}\right]\nonumber\\
&=&2\rho\left(\frac{1}{2M}\right)\cdot\left[\frac{1}{M}\sum_{i=0}^{M-1}
Q\left(\frac{\|f_n(u_{i+1})-f_n(u_i)\|}{2\sigma}\right)-
e^{-\lambda
n}\right]\nonumber\\
&\gee&2\rho\left(\frac{1}{2M}\right)\cdot\left[Q\left(\frac{1}{2\sigma
M}\sum_{i=0}^{M-1}\|f_n(u_{i+1})-f_n(u_i)\|\right)-e^{-\lambda
n}\right]\nonumber\\
&\gef&2\rho\left(\frac{1}{2M}\right)\cdot\left(Q\left[\frac{L(f_n)}
{2\sigma M}\right]-e^{-\lambda
n}\right),
\end{eqnarray}
where the labeled inequalities are explained as follows:
(a) is due to the assumed symmetry of $\rho(\cdot)$;
(b) is by its assumed convexity; (c) follows from the union bound and by identifying the first
integral of the preceding line as the error probability of the ML decision
rule in distinguishing between the two equiprobable hypotheses: 
$\calH_0:~\bY=f_n(u_i)+\bZ$ and 
$\calH_1:~\bY=f_n(u_{i+1})+\bZ$, and by using the fact that this error
probability is given by $Q(\|f_n(u_{i+1})-f_n(u_i)\|/2\sigma)$, where $Q(s)=
\frac{1}{\sqrt{2\pi}}\int_s^\infty e^{-t^2/2}\mbox{d}t$;
(d) is by the outage constraint (\ref{constraint});
(e) is due to the convexity of the function $Q(s)$ for $s\ge 0$ (which can
easily be verified from its second derivative), and finally,
(f) is because $Q(\cdot)$ is monotonically decreasing and because $\sum_i
\|f_n(u_{i+1})-f_n(u_i)\|\le L(f_n)$, which in turn follows from the fact that 
the Euclidean norm is a metric (and
so, a straight line between any two points is always shorter than any other
curve connecting them). 

Our next step would to derive an upper bound on $L(f_n)$. 
As a preparatory step toward this end, we present the following consideration.
For the given $u\in[0,1]$, let us represent
the noise vector $\bz$ as $(z_0,\bz^\prime)$, where $z_0$ is
the component of $\bz$ in the direction of
$\mbox{d}f_n(u)$ and
$\bz^\prime=(z_1,\ldots,z_{n-1})$ is the orthogonal component of $\bz$.
From the outage constraint (\ref{constraint}),
we have that
\begin{eqnarray}
e^{-\lambda n}&\ge&\sup_{u\in[0,1]}
\mbox{Pr}\{\calO_n(u)\}\\
&=&\sup_{u\in[0,1]}\int_{\calO_n(u)}p(\bz)\mbox{d}\bz\\
&\gexe&\sup_{u\in[0,1]}\int_{-\infty}^{+\infty}p(z_0)\mbox{d}z_0
\int_{\calO_n(u,z_0)}p(\bz^\prime)\mbox{d}\bz^\prime,
\end{eqnarray}
where we have defined
$\calO_n(u,z_0)\dfn\{\bz^\prime:~(z_0,\bz^\prime)\in\calO_n(u)\}$.
By Assumption A.3, $\calO_n(u,0)\subseteq\calO_n(u,z_0)$,
for every real $z_0$, and so, we can continue the above chain of inequalities
as follows:
\begin{eqnarray}
e^{-\lambda n}&\ge&\sup_{x\in[0,1]}\int_{-\infty}^{+\infty}p(z_0)\mbox{d}z_0
\int_{\calO_n(u,0)}p(\bz^\prime)\mbox{d}\bz^\prime\\
&=&\sup_{u\in[0,1]}\int_{\calO_n(u,0)}p(\bz^\prime)\mbox{d}\bz^\prime\nonumber\\
&=&\sup_{u\in[0,1]}\mbox{Pr}\{\calO_n(u,0)\}.
\end{eqnarray}
Thus, for every $u\in[0,1]$, $\calO_n(u,0)$ must satisfy the same outage
constraint
as $\calO_n(u)$, with exponent $\lambda$. Now, due to the Neyman--Pearson
theorem and due to the Gaussianity of the noise vector,
the minimum volume of $\calO_n^{\mbox{\tiny c}}(u,0)$, subject to the
constraint, $\mbox{Pr}\{\calO_n(u,0)\}\le e^{-\lambda n}$,
is attained when $\calO_n^{\mbox{\tiny c}}(u,0)$ is an $(n-1)$--dimensional
Euclidean sphere centered at the origin.\footnote{This argument follows also
from one of
the iso--perimetric inequalities, see, e.g., \cite[Theorem 7.1.1, eq.\
(7.12d)]{Zamir14}.}
Using the Chernoff bound (which is well--known to be 
exponentially tight) and the given outage
constraint,
the radius of this sphere is easily found to be
$\sigma\sqrt{n[1+w(\lambda)]}$.

We next invoke a ``tube--packing'' argument
in the spirit of \cite[pp.\ 672--673]{WJ65} (see also 
\cite[Subsection 2.3.1]{Floor08}, \cite{KR07}). 
To comply with the outage constraint, the signal locus curve must have the
property that $\calO_n^{\mbox{\tiny c}}(u,0)$ and $\calO_n^{\mbox{\tiny
c}}(u^\prime,0)$ are disjoint whenever the $u$ and $u^\prime$ are far apart
in the sense of being points that belong to different folds of the signal locus curve. This is
because no point in space can possibly belong simultaneously to the non--outage
regions of two remote values of the parameter. In mathematical terms, this
means that in the high SNR limit,\footnote{Here the high SNR limit means 
that the volumes of $\calO_n^{\mbox{\tiny
c}}(u,0)$ are relatively very small: in particular, they 
shrink when $\sigma$ is small, and $\sigma$ is assumed very small compared to
the radius of curvature of the signal locus curve.} the volume of the body
$\calT\dfn\{\by=f_n(u)+(0,\bz^\prime):~u\in[0,1],~\bz^\prime\in\calO_n^{\mbox{\tiny
c}}(u,0)\}$ is of the exponential order of
$\int_{f_n(0)}^{f_n(1)}\|\mbox{d}f_n(u)\|\cdot\mbox{Vol}\{\calO_n^{\mbox{\tiny
c}}(u,0)\}$. We next derive an upper bound and a lower bound to
$\mbox{Vol}\{\calT\}$, so that by comparing these two bounds, we can find an
upper bound to $L(f_n)$. 

As for the upper bound to $\mbox{Vol}\{\calT\}$, due to the power constraint, 
it cannot exceed the volume of a sphere whose radius is
upper bounded by $\sqrt{nP}+r_{\mbox{\tiny cov}}\{\calO_n^{\mbox{\tiny
c}}(u)\}$, where $r_{\mbox{\tiny cov}}\{\calO_n^{\mbox{\tiny c}}(u)\}$ is
the covering radius of $\calO_n^{\mbox{\tiny c}}(u)$, which in the high SNR
limit, is negligible compared to $\sqrt{nP}$ (due to Assumption A.3), that is,
$\sqrt{n[P+o(\gamma)]}$, where $o(\gamma)$ is a term that tends to zero as
$\gamma\to\infty$. Thus,
\begin{equation}
\mbox{Vol}\{\calT\}\lexe (2\pi e[P+o(\gamma)])^{n/2},
\end{equation}
where we have used the fact that the volume of an $n$--dimensional sphere of
radius $\sqrt{nS}$ is of the exponential order of $(2\pi e S)^{n/2}$ (see,
e.g., \cite[eq.\ (7.30)]{Zamir14}).

For the lower bound to $\mbox{Vol}\{\calT\}$, we have by the aforementioned
Neyman--Pearson/iso--perimetric consideration and by the outage constraint,
\begin{equation}
\label{beforecomment}
\mbox{Vol}\{\calT\}\gexe L(f_n)\cdot\min_u\mbox{Vol}\{\calO_n^{\mbox{\tiny
c}}(u,0)\}\gexe L(f_n)\cdot(2\pi e\sigma^2[1+w(\lambda)])^{(n-1)/2}.
\end{equation}

Before we proceed, an important comment is now in order: 
the r.h.s.\ is of the same exponential order as the volume of the union of
$n$--dimensional spheres of radius $\sigma\sqrt{n[1+w(\lambda)]}$ 
(as opposed to the $(n-1)$--dimensional spheres described
above), centered at $f_n(u)$, where the union runs from $u=0$ to $u=1$.
This means that in the high SNR limit, the optimal choice of $\calO_n(u)$ is the
complement of a sphere of radius $\sigma\sqrt{n[1+w(\lambda)]}$,
independently of $u$. This is observation will be important for the
achievability part (Section IV).

Comparing the upper and the lower bounds to $\mbox{Vol}\{\calT\}$,
we obtain
\begin{eqnarray}
L(f_n)&\lexe&\frac{(2\pi e[P+o(\gamma)])^{n/2}}{(2\pi
e\sigma^2[1+w(\lambda)])^{(n-1)/2}}\nonumber\\
&\exe&
\sigma\cdot\exp\left\{\frac{n}{2}\ln\frac{P+o(\gamma)}
{\sigma^2[1+w(\lambda)]}\right\}\nonumber\\
&=&\sigma e^{n[R(\lambda,\gamma)+o(\gamma)]}
\dfn L_n^*.
\end{eqnarray}
Returning to (\ref{lowerbound}), and choosing $M=M_n=L_n^*/(2\sigma s)$ for some
arbitrary constant\footnote{In principle, the lower bound can be maximized over $s$,
but such a maximization will not affect the exponential order of the 
bound, at least not in the important case of $\rho(t)=|t|^\alpha$.}
$s > 0$, we have
\begin{eqnarray}
\sup_{u\in[0,1]}\bE\left\{\rho(g_n[\bY]-u)\bigg|\calO_n^{\mbox{\tiny
c}}(u)\right\}&\gexe&
\rho\left(\frac{s}{L_n^*}\right)\cdot[Q(s)
-e^{-\lambda n}]\nonumber\\
&\exe&\exp\{-n[E_{\mbox{\tiny U}}(\lambda,\gamma)+o(\gamma)]\},
\end{eqnarray}
completing the proof of Theorem 1 for the case where $f_n(\cdot)$ is
continuous.

Relaxing now the continuity assumption, 
and referring to Assumption A.2, we may allow discontinuities of
$f_n(\cdot)$,
even at majority of the sub-intervals $[0,1/M_n), [1/M_n,2/M_n),\ldots,
[(M_n-1)/M_n,1]$, provided that the number $M_n^{\mbox{\tiny c}}$ of 
intervals of continuity is still of the exponential order of $M_n$ defined
above. In such a case, the modification of the above derivation would have the
following ingredients:
(i) the summation over $i$ in the first steps of (\ref{lowerbound}) should
be further lower bounded by excluding terms, indexed by $i$, for which there are
discontinuities of $f_n(u)$ in the interval $i/M_n\le u < (i+1)/M_n)$; 
(ii) the application of the Jensen
inequality to the function $Q$ (step (e) in (\ref{lowerbound})) 
would be limited to the remaining terms; (iii) consequently, the
lower bound above would be multiplied by a factor of $M_n^{\mbox{\tiny c}}/M_n^*$
(which will not affect the exponent, thanks to Assumption A.2), and (iv)
$L(f_n)$, in the 
argument of the $Q$-function, would mean the sum of lengths of the
continuous parts of the signal locus curve. The tube packing argument will continue
to apply to the total volume of the disjoint union all tubes 
formed by the separate continuous parts of the signal locus curve.
This completes the proof of Theorem 1.


\section*{IV. Achievability}

Upon careful inspection of the proof of Theorem 1, one may find some
suggestive indications what would be a good achievability scheme.

First, let us look at the last line of eq.\ (\ref{lowerbound}): it is natural to
think of the term $\rho(1/2M)$ as one that can be achieved by uniform quantization:
if we quantize the unit interval uniformly and create a grid of $M$ points
with equal spacings of $1/M$ in between, then upon quantizing an arbitrary $u\in[0,1]$
into the nearest grid point, the absolute value of the quantization error
cannot exceed $1/2M$, and so, the error cost will never be above $\rho(1/2M)$. 
Thus, if we can come up with an achievability scheme for which the weak--noise error is
exactly this quantization error, we can achieve the weak--noise error bound up to
a constant factor (given by the other multiplicative terms in the last line of
(\ref{lowerbound}). But once we have quantized $u$ into one out of $M$
quantization points, it only remains to map each one of 
these points into a corresponding channel
input vector, namely, to apply channel coding. The error event of this
channel code will then have to be the outage event. But according to the
comment that follows eq.\ (\ref{beforecomment}), the best choice of the outage
event (at least in the high SNR limit) is the complement of a sphere,
independently of the transmitted code vector. It is well known that for good
lattice codes in high dimension, the Voronoi cells become closer and closer to
spheres \cite[Chap.\ 7]{Zamir14}, and so, the error event (which is
independent of the transmitted code vector) is roughly the event
that the noise vector $\bZ$ would fall outside a sphere. The outage exponent
$\lambda$ must then be
the error exponent of such a lattice code. The coding rate $R=\frac{1}{n}\ln
M$ would then be a free parameter that controls the trade-off between the
weak--noise (quantization) error and the outage (decoding error) exponent.

More precisely, let $M=e^{nR}$
be a positive integer, where $R\ge 0$ is a design parameter to be
selected later.
Consider the uniform quantization of the the unit interval
into $M$ non--overlapping bins, each of size
$\Delta=\frac{1}{M}=e^{-nR}$, and let $q(u)$ be the midpoint of the bin $i(u)$ to
which $u$ belongs, i.e., 
\begin{eqnarray}
i(u)&=&\lfloor u/\Delta \rfloor\\
q(u)&=&\Delta
\left[i(u)+\frac{1}{2}\right].
\end{eqnarray}
Let us define the modulator according to
\begin{equation}
f_n(u)=\bx[i(u)],
\end{equation}
where $\{\bx[0],\bx[1],\ldots,\bx[M-1]\}$ is (part of) a channel lattice code whose
members all lie within the sphere of radius $\sqrt{nP}$ around the origin, so
as to comply with the power constraint. 
At the receiver side, we first decode $i(u)$ from the
channel output, and then $u$ is
estimated according to
\begin{equation}
g_n[\by]=\Delta\left[\widehat{i(u)}+\frac{1}{2}\right], 
\end{equation}
where $\widehat{i(u)}$ is the decoded version of $i(u)$ (which is, of course,
a function of $\by$ only). Thus, if $i(u)$ is decoded
correctly, the (weak--noise) estimation error is simply the quantization
error, whose cost is of the exponential order of
\begin{equation}
\bE\left\{\rho(g_n[\bY]-u)\bigg|\calO_n^{\mbox{\tiny c}}(u)\right\}
\le\rho(e^{-nR})\exe e^{-n\zeta(R)}. 
\end{equation}
Referring to converse bound in Theorem 1
(see eq.\ (\ref{eu})), it is suggestive that,
in the quest for asymptotically optimal performance,
the coding rate $R$ of the lattice code would be set to
\begin{equation}
R=R(\lambda,\gamma)=\frac{1}{2}\ln\frac{\gamma}{1+w(\lambda)}=
\frac{1}{2}\ln\gamma-\frac{1}{2}\ln[1+w(\lambda)].
\end{equation}
The first term on the right--most side is obviously identified as the high--SNR
approximation to the channel capacity, $C=\frac{1}{2}\ln(1+\gamma)$, and the
second term is interpreted as the inevitable gap to capacity that must be suffered in order to
comply with the outage constraint. 

It now remains to examine the achievable error exponents of lattice codes at
rate $R(\lambda,\gamma)$ and to compare to $\lambda$. 
According to
Zamir \cite[Theorem 13.4.1]{Zamir14}, the random coding exponent
of lattice codes (w.r.t.\ the MHS ensemble) is given by
$E_{\mbox{\tiny r}}(\mu/2\pi e)$, where
\begin{equation}
E_{\mbox{\tiny r}}(x)\dfn\left\{\begin{array}{ll}
0 & x\le 1\\
\frac{1}{2}(x-\ln x-1) & 1\le x \le 2\\
\frac{1}{2}\ln\frac{ex}{4} & x\ge 2\\
\end{array}\right.
\end{equation}
and $\mu$ is the normalized volume--to--noise ratio (NVNR), 
a constant that depends on the lattice
$\Lambda$ and the noise variance $\sigma^2$, defined as
\begin{equation}
\mu(\Lambda, \sigma^2(P_{\mbox{\tiny e}}))=
\frac{[V(\Lambda)]^{2/n}}{\sigma^2(P_{\mbox{\tiny e}})},
\end{equation}
where $V(\Lambda)$ is the volume of the Voronoi cell of the lattice,
and $\sigma^2(P_{\mbox{\tiny e}})$ is the value of the noise variance such
that $\mbox{Pr}\{\bZ~\mbox{falls outside the Voronoi cell}\}$ is equal to a prescribed
value of $P_{\mbox{\tiny e}}\in(0,1)$.
For such a fixed $P_{\mbox{\tiny e}}$, we have
$\mu(\Lambda, \sigma^2(P_{\mbox{\tiny e}}))> 2\pi e$, 
but the lower bound of $2\pi e$ can
be approached arbitrarily closely by some sequence of lattices \cite[Theorem
7.7.1]{Zamir14}.
Now, if one enlarges those lattice cells by a
factor of $\sqrt{1+w(\lambda)}$ in each dimension 
(by means of reducing the rate $R\approx\frac{1}{2}\ln[(2\pi eP)/V^{2/n}(\Lambda)]$
from $C$ to $R(\lambda,\gamma)$), 
keeping the noise variance intact, this would
increase $\mu$ by a factor of $1+w(\lambda)$, and then
the error exponent would be $E_{\mbox{\tiny
r}}[1+w(\lambda)]$, which for $w(\lambda)\in[0,1]$ (or equivalently,
$\lambda\in[0,\frac{1}{2}\ln\frac{e}{2}]$), gives
\begin{equation}
E_{\mbox{\tiny
r}}[1+w(\lambda)]=\frac{1}{2}\left\{1+w(\lambda)-\ln[1+w(\lambda)]-1\right\}=
\frac{1}{2}\cdot 2\lambda=\lambda,
\end{equation}
where the second equality is
by the definition of the function $w(\lambda)$ (see eq.\ (\ref{wlambda})). Thus,
the outage constraint is met in the range where $w(\lambda)\in[0,1]$, and the
converse bound is asymptotically achieved. 

In \cite[Theorem 13.7.1]{Zamir14}, an expurgated error exponent for lattice codes is
presented, which is given by $E_{\mbox{\tiny x}}(\mu/2\pi e)$, where
\begin{equation}
E_{\mbox{\tiny x}}(x)=\left\{\begin{array}{ll}
0 & x\le \frac{4}{e}\\
\frac{1}{2}\ln\frac{ex}{4} & \frac{4}{e}\le x \le 4\\
\frac{x}{8} & x\ge 4\end{array}\right.
\end{equation}
The best achievable exponent is then obtained by 
combining these two error exponents, which results in the {\it Poltyrev error
exponent}, $E_{\mbox{\tiny P}}(x)$ \cite[Theorem 13.7.1 and subsequent
discussion]{Zamir14}.

For a general $\lambda$, let us define
$E_{\mbox{\tiny P}}^{-1}(\cdot)$ as the inverse function of
$E_{\mbox{\tiny P}}(\cdot)$. In order to comply with the outage constraint,
one must have $\mu=2\pi e E_{\mbox{\tiny P}}^{-1}(\lambda)$, which means
a coding rate whose high--SNR approximation is given by
\begin{eqnarray}
R&\approx&\frac{1}{2}\ln\frac{2\pi e P}{V^{2/n}(\Lambda)}\nonumber\\
&=&\frac{1}{2}\ln\frac{2\pi e P}{\mu\sigma^2}\nonumber\\
&=&\frac{1}{2}\ln\frac{2\pi e P}{2\pi e\sigma^2 E_{\mbox{\tiny
P}}^{-1}(\lambda)}\nonumber\\
&=&\frac{1}{2}\ln\gamma-\frac{1}{2}\ln E_{\mbox{\tiny P}}^{-1}(\lambda).
\end{eqnarray}
Thus, for a general value of $\lambda$, the weak--noise error cost exponent
of the proposed communication system is given by
\begin{equation}
\label{el}
E_{\mbox{\tiny
L}}(\lambda,\gamma)=\zeta\left[\frac{1}{2}\ln\gamma-\frac{1}{2}\ln
E_{\mbox{\tiny P}}^{-1}(\lambda)\right].
\end{equation}

Strictly speaking, the above described
quantization--and--coding scheme does not comply with
Assumption A.2 since $f_n(u)$ contains discontinuities at {\it all} quantization
boundaries. However, a small formal modification makes it
compliant with Assumption A.2: let
every two consecutive points, of the form $(u_{2i},u_{2i+1})$, be mapped to
the same codeword, so that along the interval between them, the signal locus is
a fixed point in $\reals^n$ (which is that codeword), and so, it is
trivially continuous. This way, $M_n^{\mbox{\tiny c}}$ is half of the
total number of intervals, and hence of the same exponential order, complying
with Assumption A.2. 
At the decoder, a random
selection would be made between the two parameter values that pertain to the
resulting decoded codeword. In this case, the quantization
error (or the weak--noise error) would not exceed $3/2M$, whose cost is
$\rho(3/2M)$, which is still
of the same exponential order.

We have just proved the following theorem.

\begin{theorem}
There exists a sequence $F$ of modulators ($f_n\in\calF_n(P)$),
a sequence $G$ of estimators, and a sequence $\calO$ of
families of outage events, such that
\begin{equation}
\liminf_{\gamma\to\infty}[\calE(F,G,\calO)-
E_{\mbox{\tiny L}}(\lambda,\gamma)]\ge 0,
\end{equation}
where $E_{\mbox{\tiny L}}(\lambda,\gamma)$ is defined as in eq.\ (\ref{el}).
Moreover, $E_{\mbox{\tiny L}}(\lambda,\gamma)=E_{\mbox{\tiny
U}}(\lambda,\gamma)$ in the range of $\lambda$ where $w(\lambda)\in[0,1]$.
\end{theorem}

Finally, we point out that there are improved coding schemes,
such as the one in
\cite[Theorem 13.8.3]{Zamir14}, where the Voronoi modulation is based on 
lattice shaping with dither and minimum mean square error
estimation prior to nearest--neighbor decoding. However, in the high--SNR
regime, considered in this work, their performance is essentially the same
(see, in particular, eq.\ (13.68) in \cite{Zamir14}).

\section*{V. Summary and Conclusion}

In this paper, we addressed the problem of modulation and estimation under
outage constraints. For the high--SNR regime, we 
have derived upper and lower bounds on the exponential
order of the weak--noise error cost for a given outage exponent. 
The bounds coincide for a certain range of outage exponents. The proposed
achievability scheme naturally employs lattice codes due to the spherical
shape of the optimal non--outage events in the space of noise vectors.

Beyond the obvious quest for closing, or at least shrinking, 
the gaps for every $\lambda$ and $\gamma$, some
interesting future directions may include extensions in ways: (i) the
case where the parameter is a vector rather than a scalar, (ii) the use of
feedback, (iii) multi-user situations like the multiple access channel and
the interference channel, (iv) non--Gaussian memoryless channels, and (v) Gaussian channels
with correlated noise components (colored noise).

\section*{Acknowledgment}

Interesting discussions with Ertem Tuncel are greatly appreciated.


\end{document}